\documentclass[12pt,a4paper]{article} 

\usepackage{graphicx}

\usepackage{epsfig}
\usepackage{subfig} 
\usepackage{cite}

\usepackage{color} 
\usepackage[dvipsnames]{xcolor}

\usepackage{amssymb}
\usepackage{amsmath}
\usepackage{authblk}
\usepackage{doi}
\usepackage{dsfont}
\usepackage{setspace}

\usepackage{verbatim}

 \usepackage{slashed}
\usepackage[normalem]{ulem} 

\usepackage{url}

\newcommand{\mr}{\mathrm}
\newcommand{\mc}{\mathcal}
\newcommand{\pol}{\mr{pol}}
\newcommand{\bbar}{\overline{B}}
\newcommand{\psn}{\mathbf{\Phi}_n}

\newcommand{\psrad}{\mathbf{\Phi}_{\mathrm{rad}}}

\newcommand{\muf}{\mu_\mr{F}}
\newcommand{\mur}{\mu_\mr{R}}
\newcommand{\xif}{\xi_\mr{F}}
\newcommand{\xir}{\xi_\mr{R}}

\newcommand{\SHERPA}{{\tt{SHERPA}}}

\newcommand{\MCATNLO}{{\tt{MC@NLO}}} 

\newcommand{\POLDIS}{{\tt{POLDIS}}}
\newcommand{\POWHEG}{{\tt{POWHEG}}}
\newcommand{\POWHEGBOX}{{\tt{POWHEG~BOX}}}

\newcommand{\POWHEGBOXR}{{\tt{POWHEG~BOX~RES}}}
\newcommand{\PYTHIA}{{\tt{PYTHIA}}}

\newcommand{\PYTHIAE}{{\tt{PYTHIA8}}}
\newcommand{\HERWIG}{{\tt{HERWIG}}}

\newcommand{\VINCIA}{{\tt{VINCIA}}}

\newcommand{\LHE}{{\tt{LHE}}}

\newcommand{\beq}{\begin{equation}}
\newcommand{\eeq}{\end{equation}}

\newcommand{\bea}{\begin{eqnarray}}
\newcommand{\eea}{\end{eqnarray}}

\newcommand{\gev}{\mr{GeV}}

\newcommand{\pjet}{p_\mr{jet}}

\newcommand{\ptmin}{p_{T, \mr{jet}}^\mathrm{min}}

\newcommand{\alphas}{\alpha_S}

\begin{document}

\begin{titlepage}

\vskip 0.8cm

\begin{center}
{\Large \bf Parton-shower effects in \\[2ex]
polarized deep inelastic scattering }

\vskip 1.cm

{\large {
{\bf Ignacio Borsa}{\footnote{\tt ignacio.borsa@itp.uni-tuebingen.de}}, 
{\bf Barbara J\"ager}{\footnote{\tt jaeger@itp.uni-tuebingen.de}}}}

\vskip 1.cm

{\it Institute for Theoretical Physics, University of T\"ubingen,
Auf~der~Morgenstelle~14, 72076~T\"ubingen, Germany 
} 

\vspace{1.cm}

{\bf Abstract}

\end{center}

\vspace{.5cm}
We present a Monte-Carlo program for the simulation of polarized deep inelastic scattering at next-to-leading order in QCD matched to parton shower programs building on an existing implementation of the unpolarized case in the \POWHEGBOX{} package. We discuss extensions of the \POWHEGBOX{} framework necessary to account for polarized initial states and validate the code by detailed comparisons to existing fixed-order results.  We then use the new tool to make predictions for the upcoming Electron Ion Collider. We find that parton-shower effects do have an impact on experimentally accessible distributions and improve the agreement with the next-to-next-to-leading order results.

\noindent

\end{titlepage}

\newpage

{\small \tableofcontents}

\section{Introduction}
Unprecedented means to decipher the internal structure of the proton will be provided by future electron-proton ($ep$) colliders, in particular the new Electron Ion Collider (EIC) whose construction is planned at Brookhaven National Laboratory (BNL)~\cite{AbdulKhalek:2021gbh,AbdulKhalek:2022hcn}. 
The EIC will probe deep inelastic scattering (DIS) reactions that are characterized by the exchange of a photon or massive weak gauge boson of high virtuality between the lepton and the partonic constituents of the proton in an $ep$ collision. 
An outstanding feature of the EIC will be its capability to operate with polarized beams. This will enable scientists to explore the collision of longitudinally polarized electrons and protons in a laboratory environment at higher energies than ever before. 

From polarized fixed-target DIS measurements at relatively low to moderate energies performed at CERN, SLAC, DESY, and JLAB it is known that only a part of the proton's spin is carried by its quark constituents~\cite{Lampe:1998eu,Aidala:2012mv}. Further spin contributions are associated with its gluon content and  orbital angular momentum. Constraints complementary to those coming from fixed-target experiments on the distribution of longitudinally polarized partons inside the proton, in particular the gluon, have been provided by polarized proton-proton scattering experiments performed at BNL's Relativistic Heavy Ion Collider (RHIC)~\cite{Aschenauer:2013woa}. 
While these previous polarized $ep$ and $pp$ measurements have established the relevance of a better understanding of the proton's spin structure~\cite{deFlorian:2014yva,Nocera:2014gqa,Zhou:2022wzm}, more precise results are expected from the EIC which will cover a wider kinematic range than past fixed-target DIS experiments~\cite{Aschenauer:2012ve,Aschenauer:2015ata,Borsa:2020lsz}.  An extra benefit of the EIC will be the possibility to access charged current (CC) DIS mediated by the exchange of a $W$~boson as well as neutral current (NC) reactions that include photon and $Z$ exchange contributions~\cite{Aschenauer:2013iia}.  

This experimental progress has triggered renewed interested also on the theoretical side. New predictions for $ep$ colliders profit 
from the experience gained with precision calculations and tool development in the context of {\em unpolarized} proton-proton collisions at the Large Hadron Collider (LHC). 
With the last high-energy electron-hadron scattering experiments being performed at the Hadron Electron Ring Accelerator (HERA) at DESY  more than 15~years ago,  
since that time  tremendous progress has been made on the theory side. This includes the development of techniques for the calculation of higher-order perturbative corrections up to the next-to-next-to-leading order (NNLO) of QCD and even higher orders for selected processes of low multiplicity, the systematic consideration of electroweak corrections, 
and the standardization of systematic prescriptions for the combination of precision calculations and multi purpose Monte-Carlo generators capable of providing parton showers and a modeling of non-perturbative effects innate to the environment of a hadron collider, such as \HERWIG~\cite{Corcella:2000bw,Bellm:2015jjp},  \PYTHIA~\cite{Sjostrand:2006za,Sjostrand:2014zea}, and \SHERPA~\cite{Gleisberg:2008ta,Sherpa:2019gpd}.  In particular the \POWHEG{}~\cite{Nason:2004rx,Frixione:2007vw} and \MCATNLO{}~\cite{Frixione:2002ik} methods for the matching of next-to-leading order~(NLO) calculations with parton showers are providing the basis of tools used for a plethora of LHC analyses within the Standard Model and beyond.  

In calculations for {\em polarized} scattering processes compared to the unpolarized case additional challenges arise from the appearance of the $\gamma_5$ matrix and the Levi-Civita tensor that are genuinely four-dimensional objects. Special care is required in their treatment in the context of dimensional regularization for perturbative calculations beyond the leading order (LO). Nonetheless, considerable progress has been made in the theoretical description of reactions with polarized beams since the HERA era.  For polarized $pp$ collisions NLO-QCD calculations have been presented for key QCD processes such as photon production~\cite{Gordon:1993qc,Gluck:1994iz,Gordon:1994ut,Frixione:1999gr}, heavy-quark pair production~\cite{Bojak:2001fx}, inclusive hadron production~\cite{deFlorian:1997zj,deFlorian:2002az,Jager:2002xm}, single jet~\cite{deFlorian:1998qp,Jager:2004jh}, dijet~\cite{deFlorian:1998qp},  jet+hadron production~\cite{deFlorian:2009fw}, and hadron+photon production~\cite{deFlorian:2010vy}.  
Even the NNLO-QCD corrections are available for the Drell-Yan process~\cite{Ravindran:2003gi} and $W$~boson production~\cite{Boughezal:2021wjw}.  NNLO corrections have also been obtained for the polarized splitting functions~\cite{Vogt:2008yw,Moch:2014sna,Moch:2015usa,Blumlein:2021ryt,Blumlein:2021enk,Blumlein:2022gpp}.
For polarized $ep$ collisions NNLO-QCD results are available for inclusive and differential DIS~\cite{Zijlstra:1993sh,Borsa:2020ulb,Borsa:2020yxh,Borsa:2022irn,Borsa:2022cap} as well as semi-inclusive DIS~\cite{Abele:2022wuy,Abele:2021nyo}, while the N$^3$LO results have been obtained for the polarized structure function $g_1$~\cite{Blumlein:2022gpp}. For some of these fixed-order perturbative calculations private computer programs exist that allow for the computation of inclusive cross sections and various observables within realistic experimental cuts.  
Public event generators for the simulation of polarized $ep$ scattering are also available. For instance, {\tt PEPSI}~\cite{Mankiewicz:1991dp} allows to include parton shower and hadronization effects, while {\tt DJANGOH}~\cite{Charchula:1994kf, Aschenauer:2013iia} is capable of computing QED and QCD radiative corrections. There is, however, no public tool for the matching of NLO-QCD hard scattering matrix elements with a parton shower that consistently accounts for the polarization of the colliding beams.

With the current work we would like to close this gap for the phenomenologically pivotal  DIS process. Building on existing work for unpolarized $ep$ collisions~\cite{Banfi:2023mhz} we have developed an implementation of  polarized DIS at NLO-QCD accuracy in the \POWHEGBOX{}~\cite{Alioli:2010xd}, a tool for the matching of NLO calculations with parton showers using the  \POWHEG{} method. Our implementation accounts for the polarization of the incoming lepton and proton at the various levels of the calculation, starting from the hard matrix elements via the relevant parton distribution functions (PDFs), and the Sudakov factor for the matching to the parton shower in the \POWHEG{}  formalism.  

Our article is organized as follows: Technical details of our implementation are described in Sec.~\ref{sec:implementation}. In Sec.~\ref{sec:pheno} we discuss phenomenological results obtained with our new tool. We conclude in Sec.~\ref{sec:conclusions}. 

\section{Details of the implementation}\label{sec:implementation}
%
%
\subsection{Polarized DIS}
Throughout this work, we consider the deep inelastic scattering of an electron off a proton,  mediated by a photon  or massive gauge boson, that is
\beq
\label{eq:dis}
e^{-}(k_e, \lambda_e)+p(k_p, \lambda_p)\rightarrow \ell(k', \lambda')+X(\{k_X,\lambda_X\})\,,
\eeq
with $k_e$ and $k_p$ representing the momenta of the incoming electron and proton, respectively, while $k'$ corresponds to the momentum of the outgoing lepton $\ell$, which can be either an electron or a neutrino depending on whether NC or CC DIS is considered. The label $\{k_X\}$ is used to denote the four-momenta of the outgoing partons, which are collectively represented by $X$. The labels $\lambda_e,\lambda',\lambda_p,\{\lambda_X\}$, on the other hand,  indicate the helicity states  of the external particles.  
The actual hard scattering involves a parton $i$ inside the proton carrying a fraction $x$ of its momentum,  
\beq
k_i=x\,k_p\,.
\eeq
The four-momentum of the boson exchanged between the incoming lepton and parton can be determined from the lepton kinematics, and is given by 
\beq
q=k_e-k'\,, 
\eeq
with virtuality $Q^2=-q^2$. The DIS variables, i.e, the inelasticity $y$ and the Bjorken variable $x_B$  can then be defined  as
\begin{equation}
    y=\frac{q\cdot k_p}{k_e\cdot k_p}=\frac{q\cdot k_i}{k_e\cdot k_i}\,,\qquad x_B=\frac{Q^2}{2 k_p\cdot q}\,.
\end{equation}
At LO, the Bjorken variable is equal to the fraction of the proton's momentum carried by the incoming quark.

For unpolarized observables, the cross section is simply given by the sum over the different helicity configurations of the process, averaged over the number of initial-state combinations. The  case of the \textit{polarized} cross section, $\Delta\sigma$, which is the main objective of the present work, is slightly more involved.  $\Delta\sigma$  is defined as the difference between the cross sections with the helicities of the incoming particles aligned and anti-aligned, that is
\beq
\label{eq:pol_xsec_def}
\Delta\sigma\equiv\frac{1}{4}[(\sigma^{++}+\sigma^{--})-(\sigma^{+-}+\sigma^{-+})],
\eeq
where $\sigma^{\lambda_p\,\lambda_e}$ represents the cross section for a specific configuration of the incoming particles' helicities. It should be noted that the polarized cross sections in Eq.~\eqref{eq:pol_xsec_def} still involve the sum over the helicities of the final-state particles, which are treated as unpolarized. 

Up to NLO, using a notation similar to that of the original \POWHEG{} paper~\cite{Frixione:2007vw}, the mean value of any observable in polarized DIS, $\mathcal{O}^{\mathrm{pol}}$, can be expressed as 
\beq
\label{eq:pol_xsec}
\begin{split}
\langle \mc{O}^\pol\rangle&=\int\,d\psn{} \, \bigg[\mc{O}_n^\pol(\psn) \, \Delta B(\psn)+ \mc{O}_n^\pol(\psn) \,  \Delta V(\psn) \\ 
&  + \int\,d\psrad\, \mc{O}_{n+1}^\pol(\psn, \psrad) \, \Delta R(\psn, \psrad) + \int dz \,  \mc{O}_{n}^\pol(\psn)\, \Delta G(\psn,z) \bigg],
\end{split}
\eeq
where $d\psn$ indicates the phase space for the underlying Born process (including the momentum fraction of the incoming parton), and $\mc{O}_n^\pol$  ($\mc{O}_{n+1}^\pol$) corresponds to the expression of the observable $\mc{O}$ evaluated for the $n$ ($n+1$) final state momenta. $\Delta B$ is the polarized Born contribution to the cross section, while $ \Delta V$ and $ \Delta R$ are the polarized virtual and real-emission corrections. The last term in Eq.~\eqref{eq:pol_xsec} is associated with the collinear factorization terms that must be introduced in the case of cross sections involving initial hadrons, to account for divergences arising from initial-state collinear radiation. The additional variable $z$ introduced in that case corresponds to the momentum fraction of the collinear parton. It should be noted that all the previous contributions, in addition to the hard scattering matrix elements squared and appropriate normalization factors,  include the polarized PDFs.  
These are defined as the difference of the distributions of a parton with its spin being aligned or anti-aligned with the hadron’s longitudinal spin direction. 
The expression in Eq.~\eqref{eq:pol_xsec} then already includes the convolution between partonic cross sections and parton distributions. 
A summation over all possible partonic channels is implicitly assumed. 
The notation of Eq.~\eqref{eq:pol_xsec} further assumes that the $(n+1)$ particle phase space for the real emission contributions factorizes into the underlying $n$-particle phase space and a radiation phase-space (which can be parametrized in terms of three variables), as
\beq
\label{eq:ps_factorization}
d\mathbf{\Phi}_{n+1}=d\mathbf{\Phi}_n\,d\mathbf{\Phi}_{\mathrm{rad}}\,.
\eeq
The parameterizations used in this work for the LO and NLO phase spaces (including the radiation phase space) are the same as in the unpolarized DIS implementation of~\cite{Banfi:2023mhz}.

As a final comment before turning to the implementation of the polarized process in the \POWHEG{} framework, it should be noted that, as opposed to the unpolarized case, where the inclusion of the $Z$-boson contribution in NC DIS leads to minimal corrections for the kinematics to be explored at the future EIC, considering the full neutral-current process is instrumental in the polarized case, as noted in Ref.~\cite{Borsa:2022cap}. The reason for this difference can be traced back to the fact that parity-violating contributions from the massive gauge boson, that are suppressed for unpolarized processes, are enhanced for polarized scattering, leading to sizable differences between the full NC and the photon-mediated cross sections, for the values of $Q^2$ that the EIC is set to explore. We will further quantify this point in Sec.~\ref{sec:pheno}. 
%
%
\subsection{Implementation in the \POWHEGBOX{}}
To discuss the modifications necessary for our implementation of polarized DIS in the \POWHEGBOX{} it is useful to review some key aspects of the \POWHEG{} formalism. In general, the  unpolarized \POWHEG{} cross section can be cast, in a simplified way, as~\cite{Frixione:2007vw}
\begin{equation}\label{eq:POWHEG_xsec}
d\sigma_{\mathrm{PWHG}}=\overline{B}(\mathbf{\Phi}_n)\,d\mathbf{\Phi}_n\,\left\{\Delta(\mathbf{\Phi}_n, p_T^{\mathrm{min}})+\Delta(\mathbf{\Phi}_n, k_T(\mathbf{\Phi}_{n+1}))\frac{R(\mathbf{\Phi}_{n+1})}{B(\mathbf{\Phi}_{n})}\right\}\,.  
\end{equation}
Similar to the notation used for a polarized observable in Eq.~\eqref{eq:pol_xsec}, here  the quantities $B(\mathbf{\Phi}_{n})$ and $R(\mathbf{\Phi}_{n+1})$ represent the Born and real-emission contributions to the unpolarized cross section and depend, respectively, on the phase spaces for $n$ and $n+1$ particles,  $\mathbf{\Phi}_{n}$ and $\mathbf{\Phi}_{n+1}$. It is assumed that the latter can be factorized in a similar fashion as in Eq.~\eqref{eq:ps_factorization}. 
The function $k_T(\mathbf{\Phi}_{n+1})$ measures the hardness of the emitted radiation and should be equal, near the infrared limits, to the transverse momentum of the emitted parton.  The parameter $p_T^{\mathrm{min}}$  corresponds to an infrared cut-off, below which  real radiation is considered to be unresolved (typically chosen to be of order 1 GeV).

The quantities $\overline{B}(\mathbf{\Phi}_n)$ and $\Delta(\mathbf{\Phi}_n, p_T)$
in Eq.~\eqref{eq:POWHEG_xsec}, on the other hand, are inherent to the \POWHEG{} formalism. 
The \POWHEG{} function $\overline{B}(\mathbf{\Phi}_n)$ is defined as
\begin{equation}\label{eq:POWHEG_btilde}
\begin{split}
\overline{B}(\mathbf{\Phi}_n)= B(\mathbf{\Phi}_n)+ V(\mathbf{\Phi}_n)&+\left[\int d\mathbf{\Phi}_{\mathrm{rad}}\,C(\mathbf{\Phi}_{n+1})+  \int dz \,G(\mathbf{\Phi}_{n},z)\right]\\ 
&+\left[\int d\mathbf{\Phi}_{\mathrm{rad}} \left[R(\mathbf{\Phi}_{n+1}) - C(\mathbf{\Phi}_{n+1})\right] \right]\,, 
\end{split}
\end{equation}
with $V(\mathbf{\Phi}_n)$ and  $G(\mathbf{\Phi}_{n}, z)$  denoting the unpolarized virtual corrections and collinear factorization terms, respectively. The remaining terms $C(\mathbf{\Phi}_{n+1})$ correspond to the infrared subtraction terms that are needed for a finite numerical evaluation of the cross section. It is possible to check that the divergent contributions arising from the first square bracket cancel against those present in the virtual part, while,  by construction, the divergent behavior of the counter-terms mimics that of the real-emission part, thus rendering the second squared bracket finite. 

The \POWHEG{} Sudakov factor is defined as
\begin{equation}\label{eq:POWHEG_sudakov}
    \Delta(\mathbf{\Phi}_n,p_T)= \exp\left\{ -\int\frac{\left[d\mathbf{\Phi}_{\mathrm{rad}} \,R(\mathbf{\Phi}_{n+1})\,\theta(k_T(\mathbf{\Phi}_{n+1})-p_T)\,\right]}{B(\mathbf{\Phi}_n)}\right\}.
\end{equation}

Having reviewed the key elements of the  \POWHEG{} framework, we can discuss its possible extension to the polarized case. Starting from Eq.~\eqref{eq:pol_xsec}, it is straightforward to show that the mean value for any polarized observable can be cast in a fashion somewhat similar  to Eq.~\eqref{eq:POWHEG_xsec}. That is, for a polarized observable $\mathcal{O}^\pol$ we can write, up to NLO,
\begin{equation}\label{eq:pol_observable_pwhg}
\begin{split}
\langle \mathcal{O}^\pol\rangle = \int \, d\mathbf{\Phi}_n \, \mathcal{O}_n^\pol(\mathbf{\Phi}_n) \,& \Delta\overline{B}(\mathbf{\Phi}_n)\times\\
&\left\{1+ \int d\mathbf{\Phi}_{\mathrm{rad}} \, \frac{ \Delta R(\mathbf{\Phi}_n,\, \mathbf{\Phi}_{\mathrm{rad}})}{\Delta B (\mathbf{\Phi}_n)} \left[\frac{ \mathcal{O}_{n+1}^\pol(\mathbf{\Phi}_n,\, \mathbf{\Phi}_{\mathrm{rad}})}{\mathcal{O}_n^\pol(\mathbf{\Phi}_n)} - 1\right]\right\} ,
\end{split}
\end{equation}
where $\Delta\overline{B}(\mathbf{\Phi}_n)$ denotes the polarized  version of the \POWHEG{} $\bbar$ function, and is given by
\begin{equation}\label{eq:POWHEG_btilde_pol}
\begin{split}
\Delta \overline{B}(\mathbf{\Phi}_n)=& \Delta B(\mathbf{\Phi}_n)+ \Delta V(\mathbf{\Phi}_n)+\left[\int d\mathbf{\Phi}_{\mathrm{rad}} \Delta C(\mathbf{\Phi}_{n+1})+  \int dz\, \Delta G(\mathbf{\Phi}_{n},z)\right]\\ 
&+\left[\int d\mathbf{\Phi}_{\mathrm{rad}} \left[\Delta R(\mathbf{\Phi}_{n+1}) - \Delta C(\mathbf{\Phi}_{n+1})\right] \right]. 
\end{split}
\end{equation}
The quantities $\Delta B$, $\Delta V$, $\Delta R$ and $\Delta G$ have already been defined in Eq.~\eqref{eq:pol_xsec}, while $\Delta C$ is used to represent the infrared counter-terms for the polarized real emission contribution. Similarly to the unpolarized case, all the divergent contributions present in Eq.~\eqref{eq:POWHEG_btilde_pol} cancel, resulting in a finite value of $\Delta \overline{B}$.

Just as in the unpolarized case, it is  possible to think of the result in  Eq.~\eqref{eq:pol_observable_pwhg} as the perturbative expansion, up to order  $\alphas$, of an all-order quantity in which the leading collinear logarithms are resummed. It is then natural to extend the definition of the \POWHEG{} cross section of Eq.~\eqref{eq:POWHEG_xsec} to the polarized case as
\begin{equation}\label{eq:POWHEG_xsec_pol}
d\Delta\sigma_{\mathrm{PWHG}}=\Delta\overline{B}(\mathbf{\Phi}_n)\,d\mathbf{\Phi}_n\,\left\{\Delta^{\mathrm{pol}}(\mathbf{\Phi}_n, p_T^{\mathrm{min}})+\Delta^{\mathrm{pol}}(\mathbf{\Phi}_n, k_T(\mathbf{\Phi}_{n+1}))\frac{\Delta R(\mathbf{\Phi}_{n+1})}{\Delta B(\mathbf{\Phi}_{n})}\right\}, 
\end{equation}
where we have introduced the polarized \POWHEG{} Sudakov factor which is defined in analogy to the unpolarized case of Eq.~\eqref{eq:POWHEG_sudakov}, replacing the respective real and Born contributions with  their polarized counterparts, $\Delta R(\mathbf{\Phi}_{n+1})$ and $\Delta B(\mathbf{\Phi}_{n})$, i.e.\ 
\begin{equation}\label{eq:pol_POWHEG_sudakov}
    \Delta^\pol(\mathbf{\Phi}_n,p_T)= \exp\left\{ -\int\frac{\left[d\mathbf{\Phi}_{\mathrm{rad}} \,\Delta R(\mathbf{\Phi}_{n+1})\,\theta(k_T(\mathbf{\Phi}_{n+1})-p_T)\,\right]}{\Delta B(\mathbf{\Phi}_n)}\right\}.
\end{equation}

As can be seen in Eq.~\eqref{eq:POWHEG_xsec_pol}, the generalization of the DIS implementation of Ref.~\cite{Banfi:2023mhz}  to account for the polarization of initial-state particles requires a consistent replacement of the  quantities, $\bbar$, $\Delta^{\mathrm{pol}}$, $R$ and $B$,  with their polarized analogs. The implementation of $\Delta B$ and $\Delta R$  requires replacing helicity-averaged expressions for partonic matrix elements and PDFs with the relevant combination of the contributions of fixed initial-state helicities.

The case of $\Delta\overline{B}$ is, however,  more involved, since in addition to the virtual contributions it requires modifying the subtraction scheme that \POWHEG{} uses to render the cross-section numerically finite. The implementation of a subtraction scheme that accounts for the polarization of the initial-state particles is one of the key modifications of our new DIS code and can, in principle, be applied to any other process with polarized initial-state particles.  Since the \POWHEGBOX{} currently applies the FKS subtraction method~\cite{Frixione:1995ms}, we perform the necessary modifications to generalize it to the polarized case. The extension of the FKS method to polarized processes is in fact not new, and was already discussed in detail in~\cite{deFlorian:1998qp, deFlorian:1999ge}. In the following we only briefly discuss 
aspects of such an extension pertinent to our DIS implementation. 

It should be noted that, since the polarized cross section only involves differences between the helicity states of the \textit{incoming} particles, it is in principle sufficient to introduce new elements of the subtraction associated with initial-state singularities. In contrast, the helicity states of all the outgoing particles are summed over and therefore treated as unpolarized, making it unnecessary to modify the subtraction terms for final-state singularities beyond the replacement of the Born matrix elements with their polarized equivalents. The  implementation of a polarized FKS subtraction scheme is further simplified by noting that the eikonal factors appearing in the soft limit of matrix elements are independent of the helicity state of the involved particles and, for that reason, the counter-terms for soft singularities only require the modification of the Born matrix element.  The remaining initial-state collinear singularities can be dealt with by replacing the unpolarized Born cross section with its polarized counterpart and making the additional substitution
\begin{equation} \label{eq:polarized_kernels}
P_{ab}^{<}(z,0)\rightarrow \Delta P_{ab}^{<}(z,0),\qquad P_{ab}^{'<}(z,0) \rightarrow \Delta P_{ab}^{'<}(z,0),
\end{equation}
where $P_{ab}^{<}(z,\epsilon)$ is the Altarelli-Parisi kernel for $z<1$ in $d=4-2\epsilon$ dimensions, and $P_{ab}^{'<}(z,0)$ corresponds to the first term in an $\epsilon$-expansion. The corresponding polarized kernels in $4-2\epsilon$ dimensions,  $\Delta P_{ab}^{<}(z,\epsilon)$, from which both terms of  Eq.~\eqref{eq:polarized_kernels} are obtained, can be found in Ref.~\cite{Vogelsang:1996im}. 
In the \POWHEGBOX{}, the modified splittings necessary for the subtractions scheme are implemented in the subroutine {\tt collisralr}, where the approximation for the initial-state real emission is obtained in the collinear limit. 

We have checked that, with the previous modifications to account for the polarization of the incoming particles, the counter-terms generated by the  \POWHEGBOX{} correctly reproduce the divergent behavior of the real emission cross-section in all the infrared divergent limits. It is straightforward to perform these tests within the \POWHEGBOX{}, as the routine {\tt checklimits}  allows to compare the behavior of the real emission contribution with its soft and collinear approximations.

The last remaining elements to be modified in order to implement the NLO polarized cross section concern the polarized virtual contributions, $\Delta V$,  and the polarized collinear factorization  terms, $\Delta G$. The case of  $\Delta V$ is straightforward since, just as with the Born and real-emission contributions, it amounts to extracting the correct linear combinations of the helicity-dependent squared-amplitudes at parton level. 
As for $\Delta G$, it is possible to check that the replacements of Eq.~\eqref{eq:polarized_kernels} are sufficient.  To see this, it is useful to notice that, similarly to the unpolarized case described in Ref.~\cite{Frixione:1995ms}, 
the collinear divergencies stemming from the factorization terms in Eq.~\eqref{eq:POWHEG_btilde_pol} cancel against those coming from the initial-state collinear counter-terms. The remaining divergent terms are associated either with the soft limit or with final-state collinear configurations, both of which are the same in the unpolarized and polarized cases. Therefore, the only necessary modification for the polarized cross section concerns the finite collinear remnants, which can be obtained from their unpolarized counterparts using the same replacements as in Eq.~\eqref{eq:polarized_kernels}. Within the \POWHEGBOX{}, the polarized Altarelli-Parisi kernels for $z<1$, necessary for the finite remnants of the collinear factorization, are implemented in the subroutine {\tt btildecoll}.

In addition to the aforementioned replacements of unpolarized quantities with their polarized counterparts, some minor modifications related to the handling of negative weights in the \POWHEGBOX{} are necessary for the correct calculation of the fixed-order and hardest-emission cross sections in the polarized case.  At variance with the unpolarized case, since polarized cross sections are defined as a difference between helicity configurations, it is perfectly possible to obtain negative cross sections even at the lowest order in perturbation theory. Similarly,  polarized PDFs, defined as differences between PDFs with helicities aligned and anti-aligned with that of the incoming hadron, can also have negative values even at LO.  For our polarized DIS implementation it is therefore necessary to modify the \POWHEGBOX{} to allow negative values of polarized PDFs, which otherwise would be discarded internally by the subroutine  {\tt genericpdf0}. The other necessary modification concerns the generation of the upper bound for radiation, for which negative Born configurations are discarded. A modified {\tt refuse\char`_pdf}  subroutine allows negative Born contributions to be considered while generating the normalization of the upper bound for radiation. 
It should be mentioned that, by default, in the \POWHEGBOXR{} the {\em absolute} value of the ratio between the real-emission and Born contributions is considered during the generation of the upper bounding functions and the application of the veto algorithm. 

With the previous modifications, a proper handling of negative-weight events is possible within the \POWHEGBOX{}, through the use of the  {\tt withnegweights} flag. This feature has been implemented  in~\cite{Alioli:2010qp} mainly to keep track of negative contributions arising from regions of the phase space where the NLO corrections become negative and larger than the LO ones, and where the stability of the perturbative expansion up to NLO is not trustworthy.  Schematically, when this strategy is employed, the code generates a fraction of positive and negative weighted events, $n_+$ and $n_-$ given by
\beq
\label{eq:event_fraction}
n_+=\frac{\sigma_{(+)}}{\sigma_{(+)}+|\sigma_{(-)}|}\qquad n_-=\frac{\sigma_{(-)}}{\sigma_{(+)}+|\sigma_{(-)}|},
\eeq
with $\sigma_{(+)}$ and $\sigma_{(-)}$ representing the positive and negative contributions to the \textit{total} cross section, respectively. The positive- and negative-weight events are then assigned fixed weights of  $\pm\left(\sigma_{(+)}+|\sigma_{(-)}|\right)$, so that the mean value of the cross section is given by
\beq
\sigma = +\left(\sigma_{(+)}+|\sigma_{(-)}|\right) n_+ - \left(\sigma_{(+)}+|\sigma_{(-)}|\right) n_- = \sigma_{(+)} - \sigma_{(-)} = \sigma_{\mathrm{NLO}},
\eeq
and thus the correct value for the NLO inclusive cross section is obtained.

As a final comment, we would like to stress that, while in principle it is possible to avoid some of the issues concerning negative contributions in polarized processes by 
calculating the cross section contributions of Eq.~\eqref{eq:pol_xsec_def} for each of the different helicity configurations separately within the \POWHEGBOX{} framework, and then taking the appropriate linear combination, technical difficulties associated with this approach  complicate its actual implementation: Separating the cross section into terms that are positive-defined at the Born level would require using fixed-helicity PDFs, which are not standard. 
Furthermore, this approach would require computing the difference between contributions that can be large and similar in size, resulting in numerically delicate cancellations. We therefore refrain from following this strategy.

\section{Phenomenological results}
\label{sec:pheno}
We consider the inclusive production of a jet in polarized electron-proton scattering, given by 
\beq
\label{eq:jet-dis}
e^{-}(k_e)+p(k_p)\rightarrow \ell(k')+\mathrm{jet}(\pjet)+X\,,
\eeq
where a notation similar to that of Eq.~\eqref{eq:dis} has been employed. Depending on whether neutral or charged-current DIS is considered, the outgoing lepton $\ell$ is either an electron or a neutrino, respectively. Once again, in the case of NC DIS we consider both photon and $Z$-boson exchange. The only difference with Eq.~\eqref{eq:dis} is the introduction of the final state jet, with $\pjet$ indicating its momenta. The final state jet is commonly characterized by its transverse momentum, $p_T$, and pseudorapidity, $\eta$. 
We analyze the process in the laboratory frame, where the jet has a non-vanishing transverse momentum already at $\mathcal{O}(\alpha_S^0)$. This is at variance with the Breit frame, where the virtual gauge boson and the incoming parton collide head-on, and a first non-vanishing contribution to the production of observable jets is obtained at $\mathcal{O}(\alpha_S)$, with two final-state partons of opposite transverse momentum.
For the remainder of the phenomenological analysis, we consider a typical EIC setup, with energies of 
\beq
\label{eq:eic-energy}
E_{e}=18~\gev \quad\text{and} \quad E_p=275~\gev\,,
\eeq 
for the electron and proton beams, respectively, imposing the cuts
\begin{equation}\label{eq:cuts_lept}
    49\,\mathrm{GeV}^2\leq Q^2 \leq 1000\,\mathrm{GeV}^2\quad\mathrm{and}\quad 0.04<y<0.95\,.
\end{equation}
The increased lower cut on $Q^2$ compared to Refs.~\cite{Borsa:2020ulb,Banfi:2023mhz} is imposed to avoid the low-$Q^2$ region where large cancellations between partonic channels take place, as will be discussed in the following subsections. 
As for the generation of underlying Born events within the \POWHEGBOX{}, a generation cut of $Q^2>49\,\mathrm{GeV}^2$ is used (it should be noted that the momentum mappings in both  \POWHEG{} and \PYTHIA{} preserve the lepton kinematics and thus the value of $Q^2$), while no suppression factor is implemented.

Jets are reconstructed using the anti-$k_T$ algorithm~\cite{Cacciari:2008gp} with $R=0.8$ and the standard $E$-recombination scheme using the {\tt FastJet}~implementation~\cite{Cacciari:2011ma}. For our analysis we require the hardest jet to satisfy 
\begin{equation}\label{eq:cuts_jet}
    5\,\mathrm{GeV}\leq p_T \quad\mathrm{and}\quad |\eta|<3\,.
\end{equation}

Furthermore, we set the values of the factorization and renormalization scales, $\muf$ and $\mur$, to 
\beq
\label{eq:scales}
\muf=\xif \mu_0\,,\quad
\mur=\xir \mu_0\,,\quad \text{with} \quad
\mu_0^2=Q^2\,,
\eeq 
and, when indicated, we estimate theoretical scale uncertainties performing a  standard 7-point variation of the scale parameters $\xir$, $\xif$ by a factor of two.  We consider a number $N_f=4$ of active quark flavors, and use the NLO set of polarized parton distributions of the DSSV collaboration~\cite{DeFlorian:2019xxt}.

Since, to the best of our knowledge, no polarized parton shower (PS) is publicly available, we resort to the standard unpolarized showers implemented in \PYTHIAE{}. It should be noted that, even using an unpolarized shower, the dominant leading logarithmic contributions to the cross section are correctly reproduced in the NLO+PS results. As mentioned previously, since in the soft limit the splitting functions for unpolarized and polarized processes coincide, the initial-state soft-collinear enhanced behavior of the cross section is correctly captured in our NLO+PS predictions. Additionally, given that the polarized Altarelli-Parisi splitting function for a $q\rightarrow q+g$ branching is equal to its unpolarized counterpart, the contributions associated with the emission of gluons from an initial-state quark line are also described correctly. Final-state radiation, on the other hand, is treated as unpolarized in our work and therefore does not require any modification of the corresponding shower. In any case, our \POWHEGBOX{} implementation can be straightforwardly interfaced to polarized parton showers once they are developed, and paves the way for future polarized NLO+PS studies.

All the following results are obtained interfacing our new event generator for polarized DIS with the standard antenna shower of \VINCIA~\cite{Brooks:2020upa},  as implemented in  \PYTHIAE{}~\cite{Bierlich:2022pfr}. Effects associated with QED radiation are, on the other hand, neglected in the results below. The matching procedure to \PYTHIAE{} is analogous to the one introduced for unpolarized DIS in Ref.~\cite{Banfi:2023mhz}.

\subsection{Code validation}
%
%
%
\begin{figure}
    \centering
    \includegraphics[width=\textwidth]{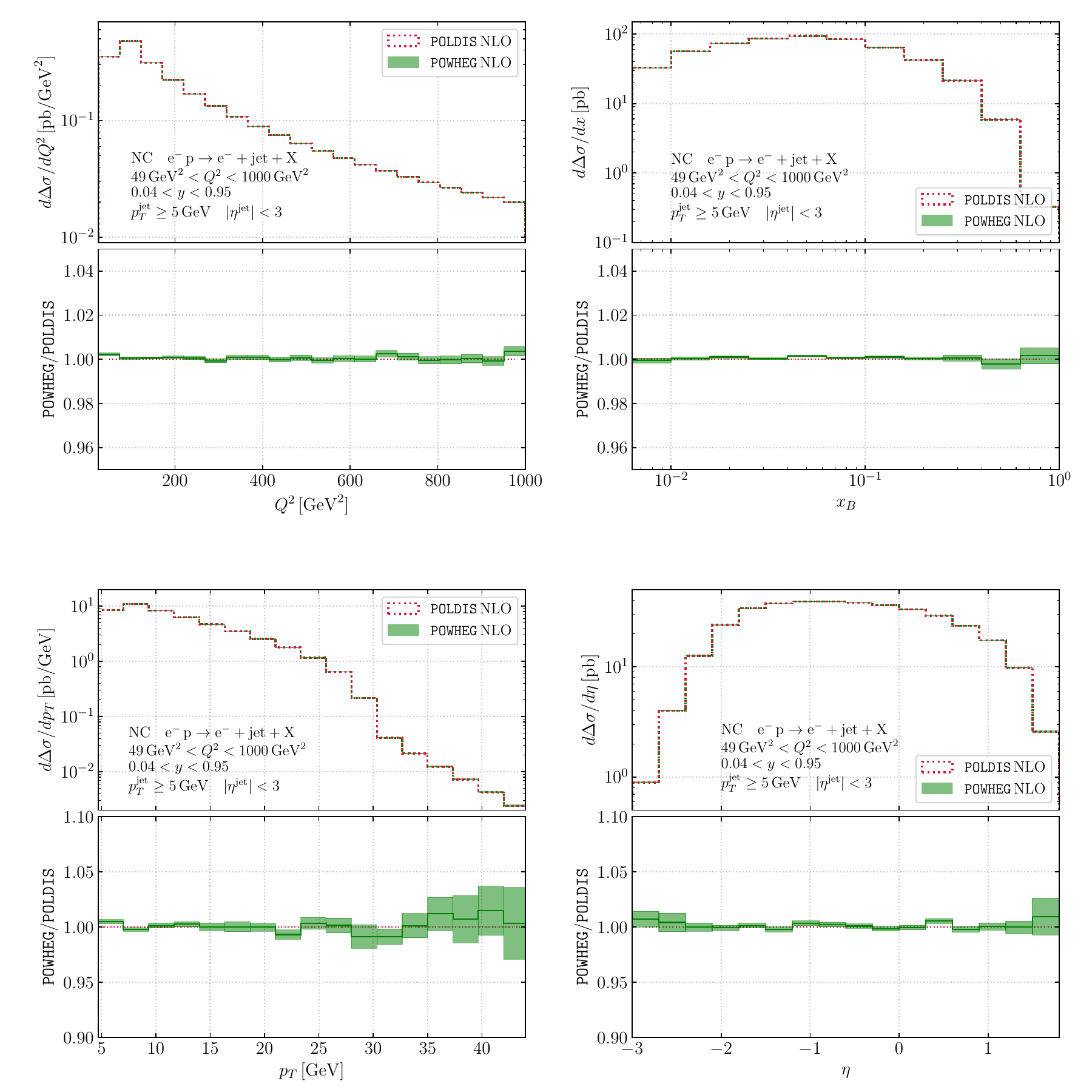}%
    \caption{
    \label{fig:NLOcomp}
    Distributions of $Q^2$ (upper left), $x_B$ (upper right), $p_T$ (lower left), and $\eta$ (lower right) for NC DIS at the EIC within the setup of  Eqs.~\eqref{eq:eic-energy}--\eqref{eq:cuts_jet} at NLO,  obtained with the \POLDIS{} (dashed red lines) and \POWHEGBOX{} (solid green lines) implementations. The bands indicate the statistical uncertainties stemming from the numerical integrations. The lower panels display the ratios of the respective \POWHEGBOX{} to the  \POLDIS{} results.
    }
\end{figure}
%
%
Before turning our attention to the phenomenological implications of matching the fixed-order calculation to a parton shower, in the following subsection we present a validation of our new  \POWHEGBOX{} implementation of polarized DIS, analyzing the events generated by \POWHEG{} for the hardest branching and performing a detailed comparison to preexisting fixed-order calculations. Additionally, we study the effects of interfacing the shower up to the partonic level, i.e., neglecting hadronization effects, and compare those effects with the NNLO corrections for the various observables. For the purpose of this validation, we show only NC DIS results, although similar findings are obtained in the case of CC.

As a first step we analyze the correct implementation of the polarized NLO cross section in the \POWHEGBOX{} framework. The fixed-order NLO cross section plays a fundamental role in the \POWHEG{} scheme, since it is used to sample the underlying Born kinematics, from which the hardest-emission events are subsequently generated. 
Therefore, we compare cross sections and some representative NLO distributions for the production of a jet in polarized DIS, as obtained with \POWHEG{}, to corresponding results obtained with the fixed-order calculation implemented in the code {\tt POLDIS}~\cite{Borsa:2020ulb, Borsa:2020yxh}. For polarized DIS cross sections within the EIC setup of  Eqs.~\eqref{eq:eic-energy}--\eqref{eq:cuts_jet} we find full agreement between the two calculations at NLO accuracy. 
In Fig.~\ref{fig:NLOcomp}, within the same setup,  we present the NLO cross section as a function of the DIS variables $Q^2$ and $x_B$, as well as of the hardest jet's transverse momentum  and its pseudorapidity, together with the respective ratios between the results of both codes. The bands in Fig.~\ref{fig:NLOcomp} correspond to the statistical uncertainty from the numerical integration of the cross section by the package {\tt MINT}~\cite{Nason:2007vt} within \POWHEG{}. For all of these distributions the results of the two programs show complete agreement within their respective uncertainties. 

Having validated the polarized NLO cross section, we now study the generation of the hardest emission in the \POWHEGBOX{}, i.e., the production of the first branching from an underlying Born configuration, according to Eq.~\eqref{eq:POWHEG_sudakov}.  The presence of the \POWHEG{} Sudakov factor in Eq.~\eqref{eq:POWHEG_sudakov}  involves, in addition to the pure NLO corrections, the resummation of contributions  associated with multiple emissions enhanced in the soft/collinear regions. The cross section for the hardest branching can be obtained from the Les Houches Events (LHE) files generated in \POWHEG{} before the application of the shower Monte Carlo. In the following, we will refer to these NLO results amended by a Sudakov factor via the \POWHEG{} procedure as \LHE{} results.

\begin{figure}
    \centering
    \includegraphics[width=\textwidth]{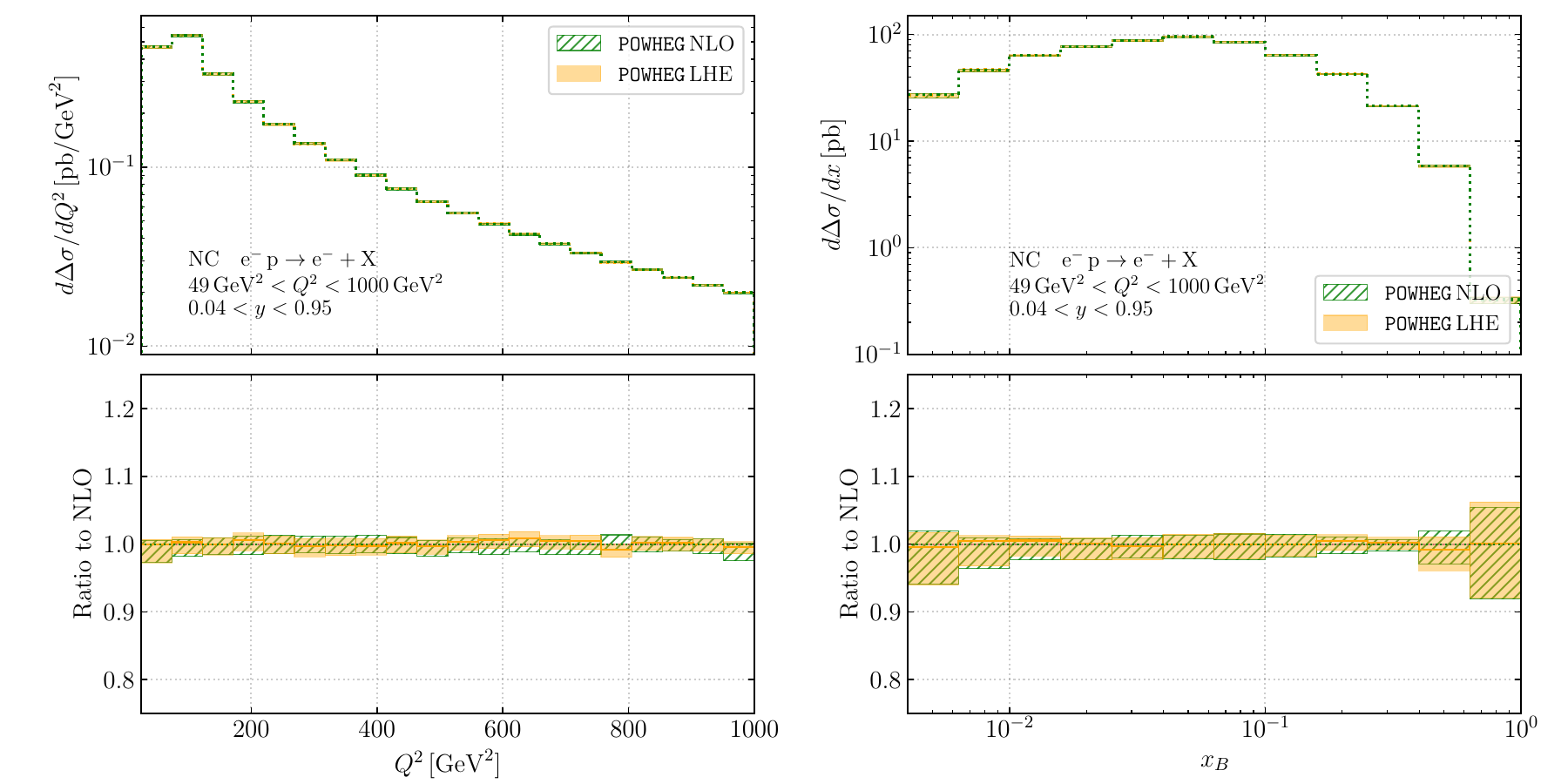}
    \caption{
        Distributions of $Q^2$ (left) and $x_B$ (right) for NC DIS  at the EIC within the setup of  Eqs.~\eqref{eq:eic-energy} and \eqref{eq:cuts_lept}  at NLO (green) and \LHE{} level (yellow) as obtained with our \POWHEGBOX{} implementation. In both cases the bands indicate theoretical scale uncertainties. The lower panels display the respective ratios  to the NLO results.
    \label{fig:inclusiveDIS}}
\end{figure}

We begin by analyzing the DIS cross section \textit{inclusive in the hardest-emission}, that is, the cross section for electron-proton scattering without any selection cuts on the partonic final state. It is important to note that, for observables that are completely inclusive in the radiation not present at  Born level, and therefore insensitive to the effects of the Sudakov factor, the fixed-order NLO and the hardest-emission cross sections are expected to agree. This is precisely the case for the DIS cross section binned in either $x_B$, $Q^2$ or $y$, if no restrictions are applied to the outgoing partons.  In that sense, the comparison of the fixed-order and \LHE{} results for this inclusive distributions allows to check that the negative contributions already present at the lowest order are being generated correctly. Figure~\ref{fig:inclusiveDIS} 
%
%
%
shows the distributions in $Q^2$ and $x_B$ for DIS imposing the selection cuts of Eq.~\eqref{eq:cuts_lept}, comparing the  \POWHEG{} fixed-order NLO and the \LHE{} results.  Additionally, we show the ratios between the latter and the NLO results obtained with \POWHEG{}. It should be noted that here, at variance with Fig.~\ref{fig:NLOcomp}, the bands represent the theoretical uncertainty estimated through the 7-point variation of the factorization and renormalization scales. For most of the bins shown in Fig.~\ref{fig:inclusiveDIS}, statistical errors are small compared to this theoretical uncertainty, and so we do not include the former. As can be seen, both results are fully consistent with each other within the uncertainties, with deviations of the central values that are typically at the percent level.

\begin{figure}
    \centering
    \includegraphics[width=\textwidth]{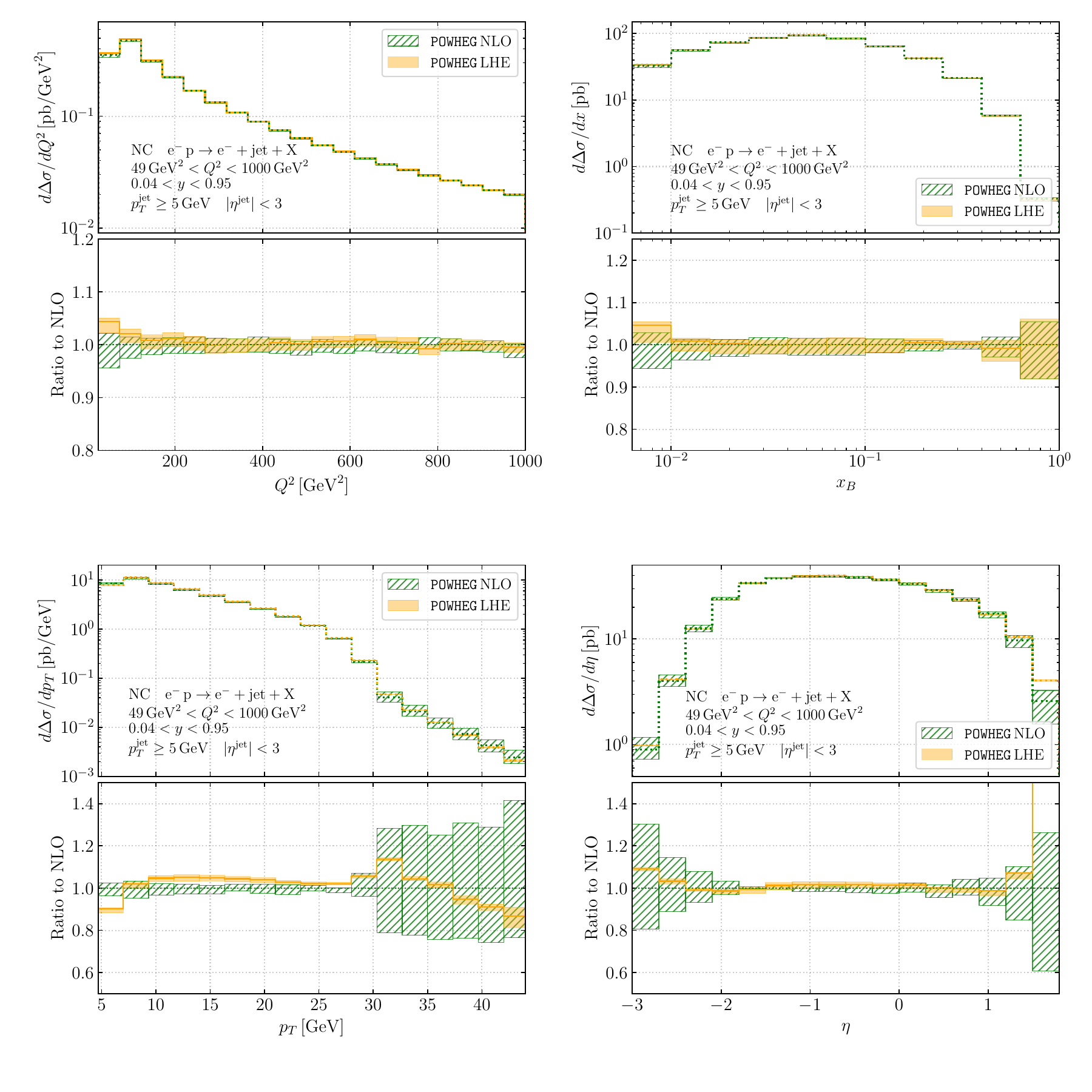}
    \caption{Similar to Fig.~\ref{fig:inclusiveDIS}, for the production of a jet in polarized DIS, using the setup of Eqs.~\eqref{eq:eic-energy}--\eqref{eq:cuts_jet}. 
    \label{fig:POWHEGcomp}
    }%
\end{figure}

After assessing the validity of the NLO and \LHE{} level results for the inclusive polarized DIS cross section, we now turn to distributions featuring an identified jet. We consider the DIS process of Eq.~\eqref{eq:jet-dis} within the cuts of Eqs.~\eqref{eq:cuts_lept} and \eqref{eq:cuts_jet}. 
Figure~\ref{fig:POWHEGcomp} 
%
%
%
%
shows the fixed-order NLO and the \LHE{} cross sections for single-jet production as a function of $Q^2$, $x_B$, as well as the transverse momentum and pseudorapidity of the leading jet. 
As in the case of the inclusive distributions of Fig.~\ref{fig:inclusiveDIS}, there is generally good agreement between the NLO and the \LHE{}  results. It should be kept in mind that, for distributions that are not completely inclusive for the emitted radiation, the distributions for the \POWHEG{} events might differ from the fixed-order NLO prediction because of the introduction of higher-order effects through the Sudakov factor. 
Unsurprisingly, the differences between the \LHE{} and NLO results for the more inclusive distributions in $Q^2$ and $x_B$ are generally small, with differences at the percent level, except for the first bin of each distribution, where the hardest-emission result exceeds the NLO one by about $4\%$. 

Differences are more pronounced in the case of the less inclusive jet distributions.  
In particular, in the transverse-momentum distribution we observe a slight damping of the low-$p_T$ region at the hardest-emission level where Sudakov effects are expected to be important. The \LHE{} result is then slightly enhanced in the intermediate $p_T$ range. At higher values of $p_T$  there is better agreement between both distributions, albeit within larger uncertainties. The rapid increase of the scale dependence in the large-$p_T$ region can be related to the fact that, at LO, the value of $p_T$ is given by $p_T^2=(1-y)Q^2$. The upper cut on $Q^2$ then implies that the region of $p_T\gtrsim 31$ GeV is not accessible at that order, and therefore the distribution is only LO accurate there.
The NLO and the \LHE{} results for the jet rapidity distributions are also  very similar, with differences of up to $2\%$ in the central region, while an enhancement of the hardest-emission result can be observed as larger values of $|\eta|$ are approached. For this last region, which is also suppressed at the LO, corrections from higher orders are again expected, as could have been anticipated by the increase of the scale-uncertainty bands.

\begin{figure}
    \centering
    \includegraphics[width=0.45\textwidth]{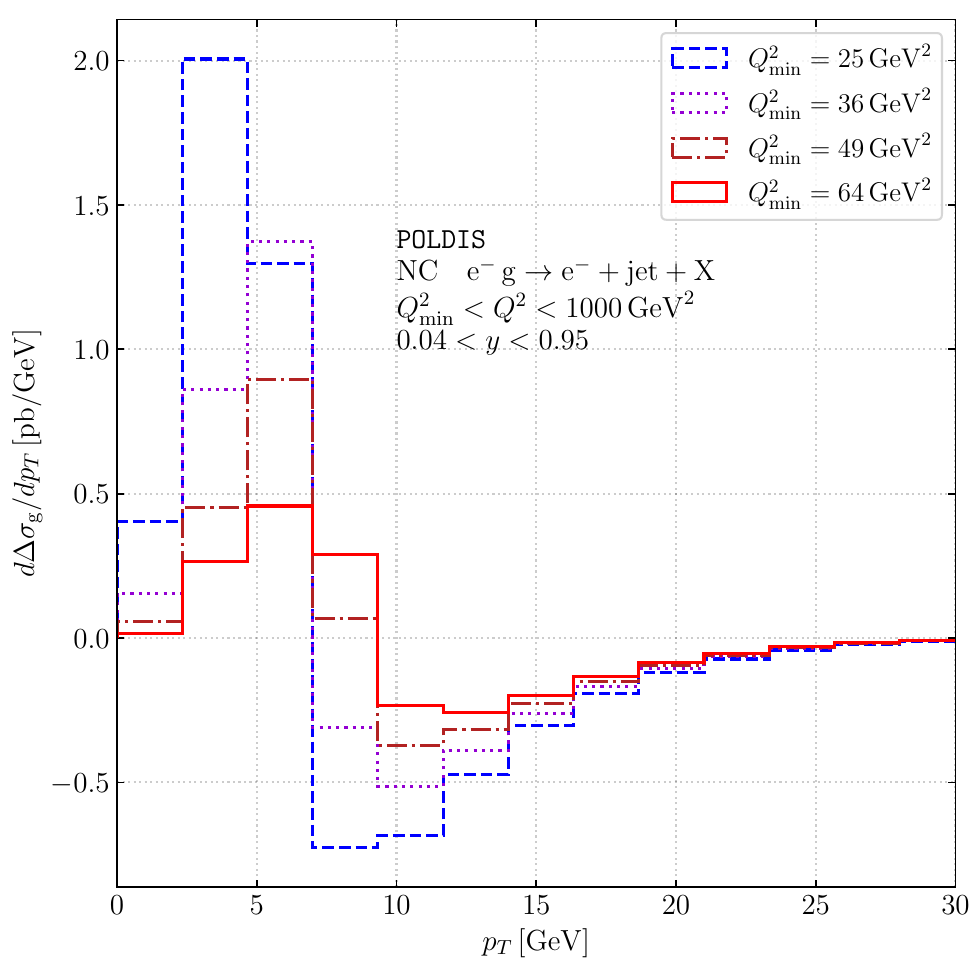}
    \caption{Real-emission contributions with initial-state gluons to the DIS cross section, for the beam energies of Eq.~\eqref{eq:eic-energy}, amended by the cuts $0.04<y<0.95$  and $Q^2_\mr{min}\leq Q^2 \leq 1000\,\mathrm{GeV}$ for different values of $Q^2_\mr{min}$. The curves are plotted as functions of the leading jet's transverse momentum. 
    \label{fig:gluons_pt}
    }%
\end{figure}

While the suppression of the cross section in the low-$p_T$ limit and the enhancement of the distributions for large $|\eta|$ can be related to higher order effects introduced through the Sudakov factor, the origin of the small differences for the first bins in $Q^2$ and $x_B$, as well as the intermediate region of $p_T$ can be traced back to the sampling strategy implemented in the \POWHEGBOX{}. It should be noted that, within the \POWHEGBOX{} framework, the fraction of events generated  for each of the different Born configurations is determined by their relative contribution to the \textit{total} NLO cross section (thus, the NLO accuracy of the integrated cross section is maintained). In the case of DIS, those fractions include the contributions from the real-emission diagrams with initial-state gluons, integrated over the phase-space of the final-state partons. In the polarized case, those contributions have a non-trivial dependence on the $p_T$ of the produced jet and, depending on the specific cuts used for jet production, there can be differences of sign between the corrections to the inclusive and exclusive cross sections. 
To clarify this point, Fig.~\ref{fig:gluons_pt}
%
%
%
%
shows the contribution from processes with initial-state gluons to the inclusive cross section (without cuts on the outgoing partons), as a function of the $p_T$ of the leading jet, for different values of the lower cut on $Q^2$. The curves were obtained with the code {\POLDIS}. As can be seen  there is a change of sign of the gluonic contribution, that is typically negative for high values of $p_T$, but reverses this behavior as lower values of $p_T$ are approached, becoming large and positive. This feature is more pronounced for lower values of the cut $Q^2_\mr{min}$.  Note that the contribution of gluon-initiated processes to the total cross section involves the integration over the whole range of $p_T$, while imposing some cut $\ptmin$ 
on the produced jet would involve keeping only the region $\ptmin<p_T$. Therefore, depending on the selection of cuts, it is possible to have a positive correction from the gluons to the total cross section and a negative correction for the cross section with cuts, potentially leading to an overestimation of the contribution of some of the Born configurations. This kind of differences are not specific to the polarized case and can, in principle, be encountered also in unpolarized processes with negative real-emission corrections. Resorting to higher values of $Q^2_{\mathrm{min}}$ and lower values of $\ptmin$, as the ones chosen in Eqs.~\eqref{eq:cuts_lept}--\eqref{eq:cuts_jet}, allows to alleviate this discrepancy, resulting in meaningful results. In that sense, it is important to note that combinations of low values of $Q^2_{\mathrm{min}}$ and high values of $\ptmin$ should be avoided. 

As a last step for the validation of our new code, we turn our attention to the effects introduced by interfacing a parton shower to the NLO calculation, to which we will refer from now on as NLO+PS. For the sake of this validation, we only consider the shower phase of \PYTHIAE{}, that is, we neglect all the hadronization and multiple-particle-interaction effects, making it possible to directly check whether the higher order effects introduced via the Sudakov factor in \PYTHIA{}  are compatible with the corrections obtained at NNLO. To estimate the latter, we resort to the numerical NNLO implementation in the code \POLDIS{}. 

\begin{figure}
    \centering
    \includegraphics[width=\textwidth]{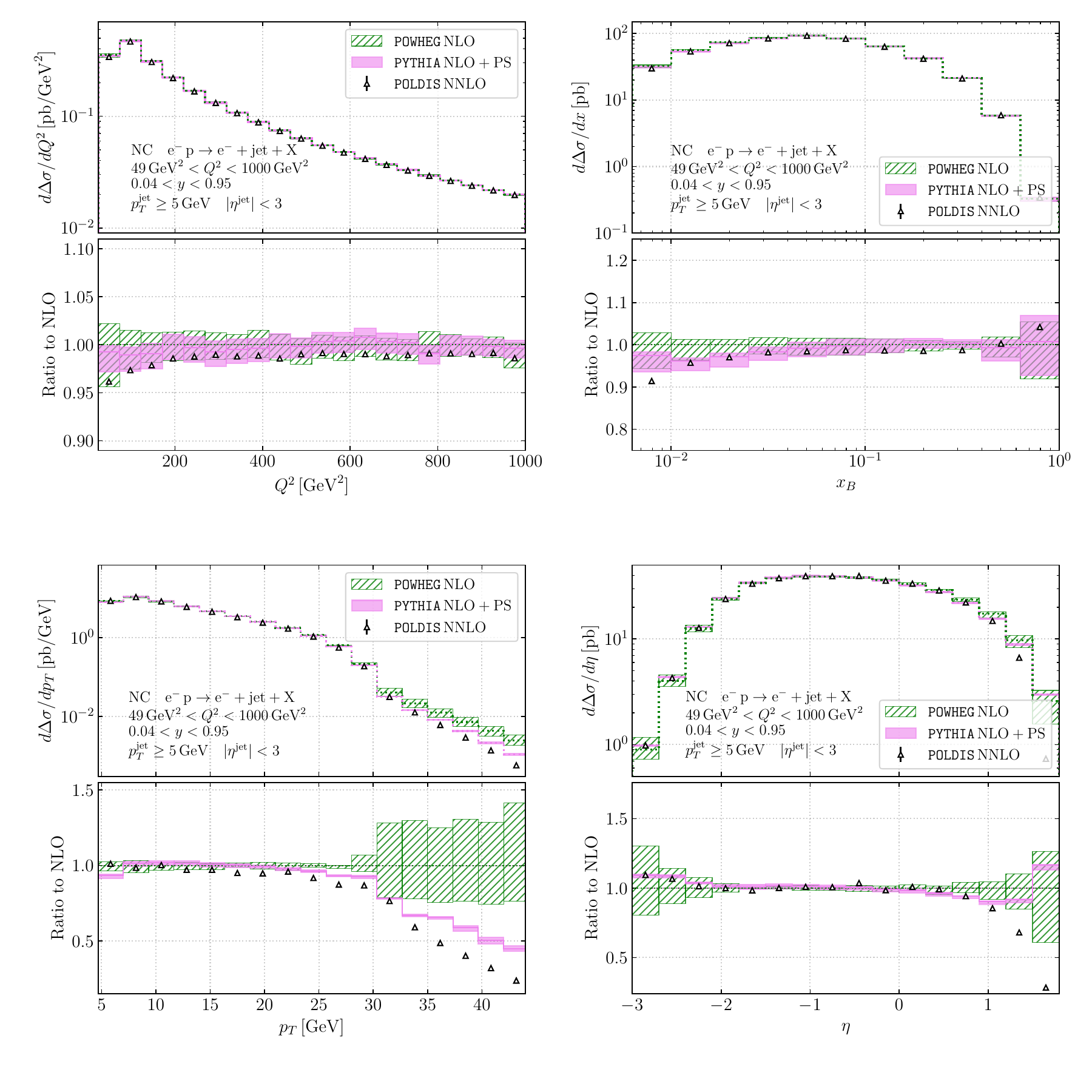}
    \caption{Distributions of $Q^2$ (upper left), $x_B$ (upper right), $p_T$ (lower left), and $\eta$ (lower right), for the production of a jet in polarized DIS, using the setup of Eqs.~\eqref{eq:eic-energy}--\eqref{eq:cuts_jet}, at NLO (green), NLO+PS (pink) and NNLO (white triangles). The NLO+PS results presented do not include hadronization or MPI effects. The lower panels show the ratios to the fixed-order NLO result. The bands indicate theoretical scale uncertainties.    
    \label{fig:PYTHIA_matching_nohad}
    }%
\end{figure}

Figure~\ref{fig:PYTHIA_matching_nohad}
%
%
%
%
presents the fixed-order NLO and the NLO+PS cross sections for single-jet production for the same distributions as in Fig.~\ref{fig:POWHEGcomp}. The figure depicts theoretical uncertainties associated with missing higher-order contributions, which are estimated by the standard 7-point variation of the renormalization and factorization scales of Eq.~\eqref{eq:scales}. To indicate the effect of the inclusion of higher-order corrections we also present the respective NNLO results. 
Just as in the case of the \LHE{} distributions, the effect of matching the PS with the NLO results is typically small for the more inclusive distributions of $Q^2$ and $x_B$. In the case of  $Q^2$, the inclusion of the PS  leads to a slight $\sim1\%$ suppression of the distribution in the intermediate region, that is accentuated as lower values of $Q^2$ are approached, reaching $2\%$ for the first bins. It is interesting to note that, for most of the intermediate and low-$Q^2$ region, this suppression agrees well with the trend shown by the NNLO corrections, and brings the NLO+PS closer to the NNLO result. For larger values of $Q^2$, the effects of the shower are negligible, with the NLO, NLO+PS, and NNLO results all agreeing within uncertainties. The effect of the parton shower on the $x_B$-distribution shows a similar trend (as expected considering that $x_B$ and $Q^2$ are correlated for DIS), with a small reduction of the cross section for low $x_B$ values, that also agrees with the suppression from the NNLO corrections.  Although not shown in Fig.~\ref{fig:PYTHIA_matching_nohad}, we have additionally checked that, for inclusive DIS, the NLO and NLO+PS results completely agree for the previous values of $Q^2$ and $x_B$.

More sizable effects are obtained for the  $p_T$ and $\eta$ distributions of the hardest jet, for which the contributions from enhanced logarithms are expected to be larger. Besides the suppression of the cross section for the first bin of $p_T$, the effect of the PS is typically small for the low-$p_T$ region. For those bins, the differences between the fixed-order and showered results are typically of order $1-2\%$ and are well contained within scale uncertainties. On the other hand, the parton shower does produce a more pronounced dampening for the intermediate region, of order $5-10\%$,  and a larger suppression for higher values of $p_T$ that can reach up to $50\%$.  As mentioned previously, this last region is kinematically forbidden at LO, and sizable effects from higher orders are expected. It is again interesting to note that, as in the case of the more inclusive distributions discussed above, the effect of the parton shower follows the general trend of the higher order corrections and greatly improves the agreement with the NNLO distributions for intermediate and large values of $p_T$. 

Similar comments can be made for the pseudorapidity distribution of the hardest jet, with the main effect of the parton shower being to shift the distribution towards lower values of $\eta$ (corresponding to the direction of the  incoming hadron), as this region gets populated by the collinear emission of partons. In this case, the inclusion of the parton shower leads to more sizable corrections in the large $|\eta|$ limit, that can reach up to $10\%$.  For most of the bins, this shift of the distribution towards the lower pseudorapidities region improves the agreement with the NNLO results as well.
%
%
\subsection{Predictions for the EIC}
Having validated our new \POWHEGBOX{} implementation we now turn to study  the phenomenology of the complete NLO+PS result for polarized DIS at the EIC. We consider the production of a jet in polarized DIS with a setup similar to that in the previous sub-section,  imposing the selection cuts of Eqs.~\eqref{eq:cuts_lept}--\eqref{eq:cuts_jet}.  

Once again, we shower the event file generated with \POWHEG{} using the antenna shower of \VINCIA~\cite{Brooks:2020upa} but, at variance with the previous section, we now take into account hadronization effects and multiple-particle-interactions (MPI). Effects associated with QED radiation are, once again, neglected in the results below. We use the Monash 2013 tune~\cite{Skands:2014pea}, which is the default in \PYTHIAE{}.  

\begin{figure}
    \centering
    \includegraphics[width=\textwidth]{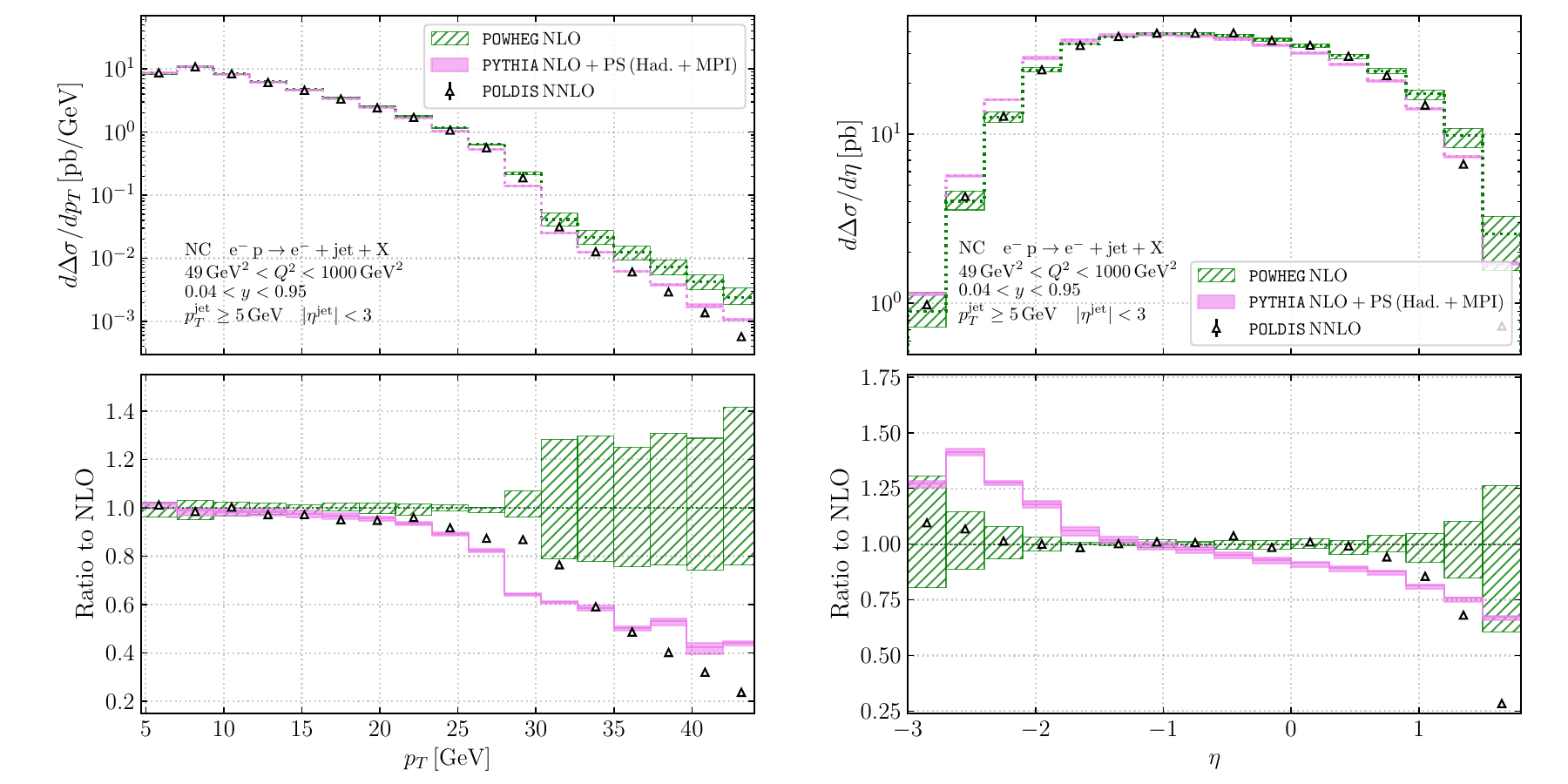}
    \caption{Similar to Fig.~\ref{fig:PYTHIA_matching_nohad}, but including hadronization and multiple-particle-interaction effects in the NLO+PS. 
    \label{fig:PYTHIA_matching}
    }%
\end{figure}

We begin analyzing the case of DIS mediated by neutral gauge bosons. Figure~\ref{fig:PYTHIA_matching}
%
%
%
%
presents a comparison between the NLO and the NLO+PS  results obtained with the \VINCIA{} shower, in a similar style to Fig.~\ref{fig:PYTHIA_matching_nohad}, for the distributions in $p_T$ and $\eta$. The distributions in $Q^2$ and $x_B$ are not included in Fig.~\ref{fig:PYTHIA_matching}, since activating the hadronization effects does not produce a significant change in those distributions besides a further suppression for low-$Q^2$.

Similarly to the distributions depicted in Fig.~\ref{fig:PYTHIA_matching_nohad}, the effect of the PS can be significant, particularly in the regions where the LO is kinematically suppressed and where the emission of extra partons is necessary to get populated. In general, the inclusion of hadronization effects does not alter the trend of the corrections induced by the parton shower, although some of its effects are enhanced. 
In the case of the transverse-momentum distribution, as compared to the results shown in Fig.~\ref{fig:PYTHIA_matching_nohad}, when hadronization and MPI effects are taken into account we observe a further suppression in the intermediate and high-$p_T$ regions, with a reduction of the cross section that can reach up to $60\%$ for the latter. 
 While the high-$p_T$ behavior of the NLO+PS is similar to the one observed for the unpolarized case in Ref.~\cite{Banfi:2023mhz}, the effects in the low-$p_T$ domain are generally small in the polarized case. 
As for the distribution in the leading jet's pseudorapidity, activating hadronization and MPI effects further increases the shift of the NLO+PS distribution towards lower values of $\eta$.  In this case, the effect of the PS can be of the order $40\%$.  
For both jet distributions, the differences between NLO and NLO+PS results are typically larger than the scale-uncertainty bands in the regions where the LO cross section is suppressed, and higher-order corrections become more relevant. For these regions, the effect induced by non-perturbative effects in \PYTHIA{} is generally larger than the NNLO corrections as well. 

\begin{figure}
    \centering
    \includegraphics[width=\textwidth]{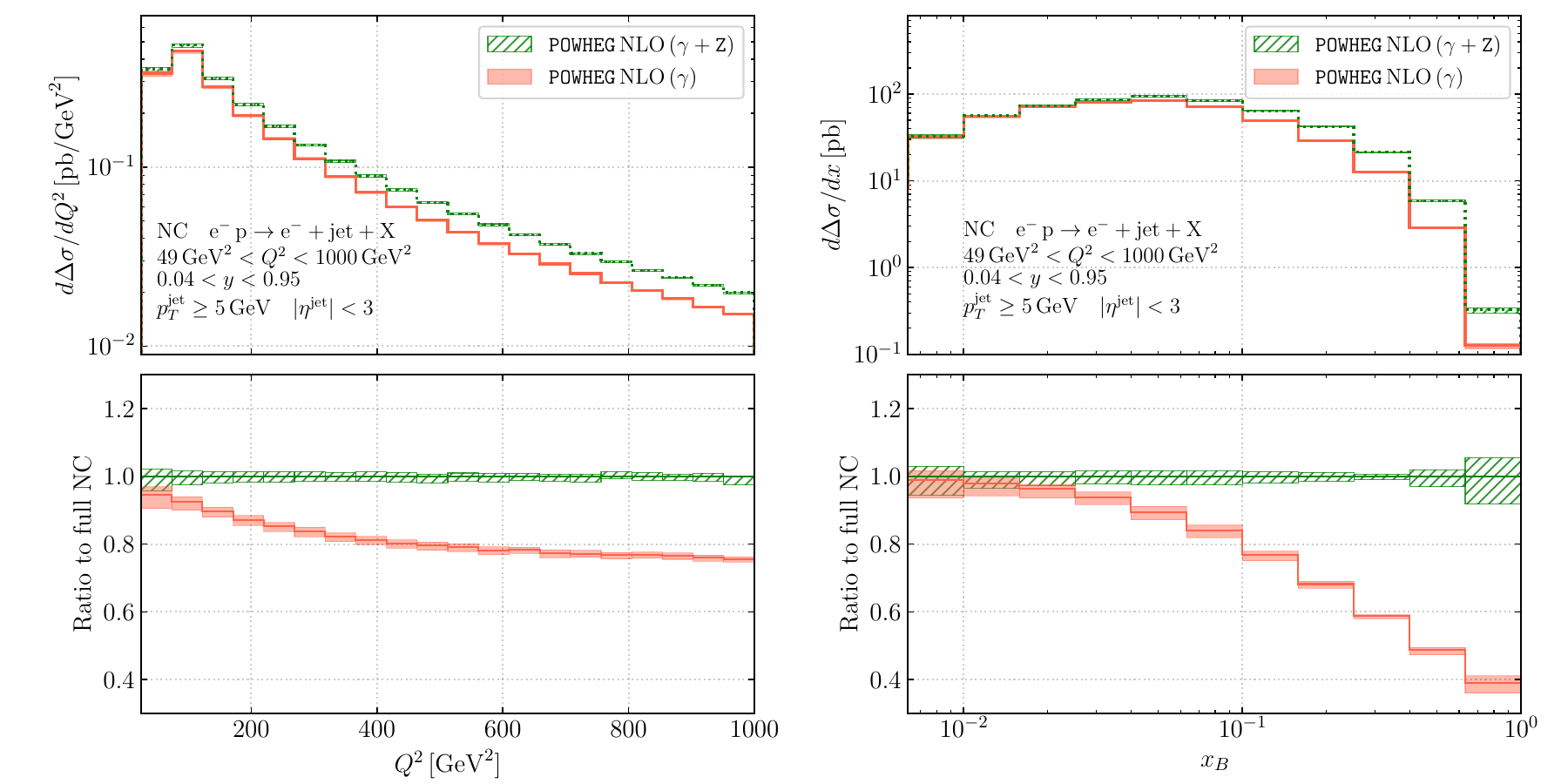}
    \caption{
Distributions of $Q^2$ (left) and $x_B$ (right) for NC DIS  at the EIC within the setup of  Eqs.~\eqref{eq:eic-energy}--\eqref{eq:cuts_jet} including both photon and $Z$-exchange (green) and only pure photon-exchange contributions (yellow) at NLO accuracy. In both cases the bands indicate theoretical scale uncertainties. The lower panels display the respective ratios to the full NC results.
     \label{fig:NC_vs_gamma}
    }%
\end{figure}

As a final remark regarding the NC results, and as mentioned in Sec.~\ref{sec:implementation}, we want to stress the importance of considering the exchange of massive $Z$~bosons (in addition to photon-exchange contributions) for a correct description of the polarized observables presented in Figs.~\ref{fig:PYTHIA_matching_nohad} and \ref{fig:PYTHIA_matching}.  To illustrate this last point, Fig.~\ref{fig:NC_vs_gamma} shows a comparison between the full NC results for the production of a jet in Eq.~\eqref{eq:jet-dis}, and analogous results considering exclusively photon-exchange. 
In both cases, the results correspond to  fixed-order NLO. 
As opposed to the unpolarized case, in which the effects stemming from $Z$ exchange are small for the probed values of $Q^2$ \cite{Borsa:2022cap}, for the polarized cross sections the differences between the full NC and the photon distribution are significant, even at lower values of $Q^2$. Such differences range from $5\%$ in the low-$Q^2$ region, where the suppression from the $Z$-boson's mass is stronger, to up to $\sim25\%$ for higher values of $Q^2$.  The difference is even more sizable in the case of the $x_B$~distribution. As noted in \cite{Borsa:2022cap}, the reason for this difference in the behavior of the unpolarized and polarized results can be traced back to the fact that parity-violating contributions, that are typically suppressed for unpolarized processes  are, on the contrary, enhanced for polarized processes.  Additionally, cancellations between the contributions of different initial-state partons, which can be large for polarized DIS with photon exchange~\cite{Borsa:2020yxh}, are not as strong for $Z$-exchange.
 %
%
%
%

\begin{figure}
    \centering
    \includegraphics[width=\textwidth]{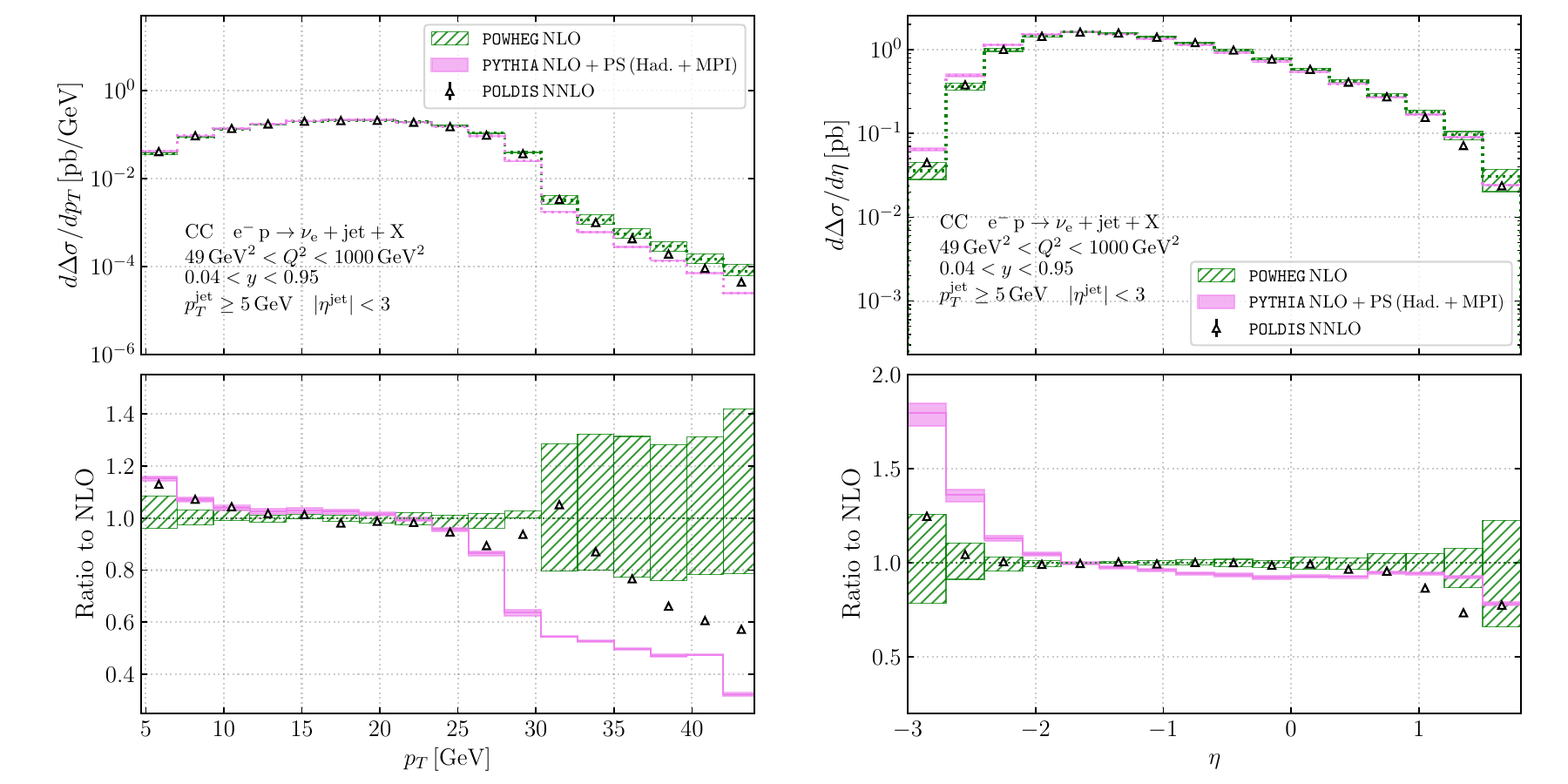}
    \caption{Similar to Fig.~\ref{fig:PYTHIA_matching}, but for charged-current DIS.
    \label{fig:PYTHIA_matching_CC}
    }%
\end{figure}

Having studied the case of neutral currents, we now discuss similar results for  charged-current DIS, in which a $W$~boson is exchanged. 
Figure~\ref{fig:PYTHIA_matching_CC} presents the analogous distributions as  Fig.~\ref{fig:PYTHIA_matching}, for DIS mediated by the exchange of a $W$-boson.
%
%
%
%
The modifications of the fixed-order NLO result by the inclusion of the parton shower generally follow the same trend as in the NC case.  For the $p_T$  distributions, the PS induces a shift towards lower values of transverse momentum, with a sizable suppression of the high-$p_T$ region. At variance with the NC case, for lower values of $p_T$ a stronger enhancement of the distribution is observed. This last feature can be explained noting that cancellations between partonic channels at low-$p_T$ are not as strong for CC DIS. The contribution to the cross section from initial $d$-quarks, which is negative for the probed values of $p_T$  in NC DIS, is suppressed for CC with an initial-state electron, because of the conservation of electric charge.   
For the $\eta$ distribution the NLO+PS results also behave similarly as the respective NC results. Once again, compared to the fixed-order NLO distribution, the NLO+PS result is shifted towards lower values of $\eta$. 
Analogously to the NC case, the corrections from the PS are typically larger than the scale-uncertainty bands for all the  regions where higher order corrections are expected to be sizable, and in general show a similar trend to that exhibited by the NNLO corrections.

\section{Conclusions and outlook}
\label{sec:conclusions}
In this work we have presented a Monte-Carlo program for the simulation of polarized DIS at NLO+PS accuracy in the framework of the \POWHEGBOX{}. Building on existing work for the unpolarized case~\cite{Banfi:2023mhz} we developed extensions of the   \POWHEGBOX{} necessary to deal with polarized initial-state particles. 
This required  the addition of  suitable subtraction terms in the context of the FKS procedure for the treatment of infrared divergences associated with polarized initial-state partons, the extension of the \POWHEG{} Sudakov  form factor to the polarized case, and a proper treatment of contributions with negative weights that, at variance with the unpolarized case, can naturally appear already at Born level because of the definition of the polarized cross section as the difference of contributions with given helicity configurations. 
Besides these general extensions of the \POWHEGBOX{} we supplied  the relevant polarized scattering amplitudes at LO, virtual corrections and real-emission contributions for the DIS process. 

After a careful validation of the developed program, we turned to phenomenological studies for the upcoming EIC. We found that depending on the considered observable and setup, PS effects can be significant, modifying fixed-order NLO results considerably in selected regions of phase space. For setups with an identified hard jet the NLO+PS results tend to be closer to the NNLO predictions than to the corresponding NLO results. 

The newly developed code will allow physicists preparing for the EIC to make use of a flexible Monte-Carlo program capable of providing predictions within realistic experimental selection cuts that can be conveniently interfaced to further tools, for instance for the simulation of detector effects, while retaining the NLO accuracy of the hard-scattering cross section. Moreover, we hope to facilitate the further development of polarized parton-shower generators in time for upcoming experiments with polarized beams.   

Our code is publicly available in the \POWHEGBOXR{} repository, see: \url{https://powhegbox.mib.infn.it}.
 
\section*{Acknowledgements}
%
We are very grateful to Giulia Zanderighi, Daniel de Florian, and Werner Vogelsang for valuable discussions. 
We would also like to thank Silvia Ferrario Ravasio for helping with the merge of the polarized version of the code.
This work has been supported by the German Research Foundation (DFG) through the Research Unit FOR 2926. 
The authors acknowledge support by the state of Baden-W\"urttemberg
through bwHPC and the German Research Foundation (DFG) through grant
no INST 39/963-1 FUGG.

\bibliographystyle{JHEP}
\bibliography{pol-dis}

\end{document}